\documentclass{aa}
\usepackage{graphicx}
\usepackage{txfonts}

\begin{document}
   \title{The first cyclotron harmonic of 4U 1538$-$52}

   \author{J. J. Rodes-Roca\inst{1,2}, J. M. Torrej\'on\inst{1},
   I. Kreykenbohm\inst{3,4}, S. Mart\'{\i}nez N\'u\~nez\inst{1}, A. Camero-Arranz\inst{5}
          \and
          G. Bernab\'eu\inst{1}
          }

   \offprints{J. J. Rodes}

   \institute{Department of Physics, Systems Engineering and Sign Theory, University of Alicante,
              03080 Alicante, Spain\\
              \email{rodes@dfists.ua.es}
         \and
             Department of Physics and Astronomy, University of Leicester, Leicester, LE1 7RH, UK
         \and
             Dr. Karl Remeis-Sternwarte, Sternwartstr. 7, D-96049 Bamberg, Germany\\
\and Erlangen Centre for Astroparticle Physics (ECAP), Erwin-Rommel-Str.~1, 91058 Erlangen, Germany\\             
         \and
             National Space Science and Technology Center, 320 Sparkman Drive, Huntsville,
             AL 35805, USA\thanks{Fundaci\'on Espa\~nola de Ciencia y Tecnolog\'{\i}a, C/ Rosario
             Pino, 14-16, 28020 Madrid, Spain}\\
                          }

   \date{Received     ; accepted       }

  \abstract
   {Cyclotron resonant scattering features are an essential tool for studying the magnetic field of neutron stars.
   The fundamental line provides a measure of the field strength, while the harmonic lines provide information
   about the structure and configuration of the magnetic field. Until now only
   a handful of sources are known to display more than one cyclotron line and only two of them
   have shown a series of harmonics.}
   {The aim of this work is to see the first harmonic cyclotron line, confirming the fundamental line at $\sim$22 keV, thus increasing
   the number of sources with detected harmonic cyclotron lines.}
   {To investigate the presence of absorption or emission lines in the spectra, we have combined
   \emph{RXTE} and \emph{INTEGRAL} spectra. We modeled the 3--100 keV continuum emission
   with a power law with an exponential cut off and look for the second absorption feature.}
   {We show evidence of an unknown cyclotron line at $\sim$47 keV (the first harmonic) in
   the phase-averaged X-ray spectra of 4U~1538$-$52. This line is detected by several telescopes at
   different epochs, even though the S/N of each individual spectrum is low.}
   {We conclude that the line-like
   absorption is a real feature, and the most straightforward interpretation is that it is
   the first harmonic, thus making 4U 1538$-$52 the fifth X-ray pulsar with more than one cyclotron line.}

   
   \authorrunning{J. J. Rodes-Roca et al.}
   \titlerunning{The first harmonic cyclotron line of 4U 1538$-$52}
   \maketitle
%

\section{Introduction}

Cyclotron resonant scattering features (CRSFs), usually referred to as
`cyclotron lines', have proved to be powerful tools for directly studying
the magnetic field in neutron stars. CRSFs are present in the hard
X-ray spectra of several X-ray pulsars and originate in the ``cyclotron
process'' under extreme conditions.  Through $E_{cyc}=11.6B_{12} \times (1+z)^{-1}$ keV
(the `12-B-12 law', where $z$ is the gravitational redshift), an energy of the fundamental feature in the hard
X-rays indicates that the magnetic fields are rather strong ($ B \sim
10^{12}$\,G). Under such conditions, the interaction of the electrons and
radiation must be treated quantum-mechanically. The behaviour of an
electron in a strong magnetic field
implies that the electron energy must be quantized in 
so-called Landau levels. These absorption features stem from the
resonant scattering of photons by electrons, also referred to as
cyclotron lines.  

While the fundamental energy of the cyclotron line provides valuable
information about the magnitude of the field, it is only through the
detection and the analysis of the harmonic lines that we can get
direct information about the geometrical configuration of the B field
(\cite{harding} 1991; \cite{araya} 2000; \cite{schonherr07} 2007).
However, to date,
only in a handful of systems have harmonic lines been discovered, and
only two systems have shown more than two (\cite{santangelo99} 1999 \&
\cite{coburn05} 2005). It is
therefore paramount to add as many systems to this
selected group as we can.

In this work, we present a spectral analysis of the high mass X-ray
binary pulsar 4U 1538$-$52. It is an eclipsing system consisting of
the B0 I supergiant star QV Nor and a neutron star with an orbital
period of $\sim$3.728 days (Clark \cite{clark}). The orbital
eccentricity is $\sim$0.08 (Corbet et al. \cite{corbet}), although more
recently a higher value of $\sim$0.17 was deduced by Clark
(\cite{clark}). The X-ray eclipse lasts $\sim$0.6 days (Becker et al.
\cite{becker}). 
The system is fairly bright in X-rays.
The estimated flux is $\sim (5-20)\times 10^{-10}$ erg s$^{-1}$ cm$^{-2}$
in the 3$-$100 keV range (Rodes~\cite{rodesPhD}). Thus, assuming a
distance of the source of $\sim$5.5 kpc (Becker
et al. \cite{becker}, Parkes et al. \cite{parkes}) and an isotropic
emission, the luminosity follows $\sim (2-7)\times 10^{36}$ erg/s.
The magnetized neutron star has a spin
period of $\sim$529 s (Davison \cite{davison}; Becker et
al. \cite{becker}).

\begin{figure*}[h!t]
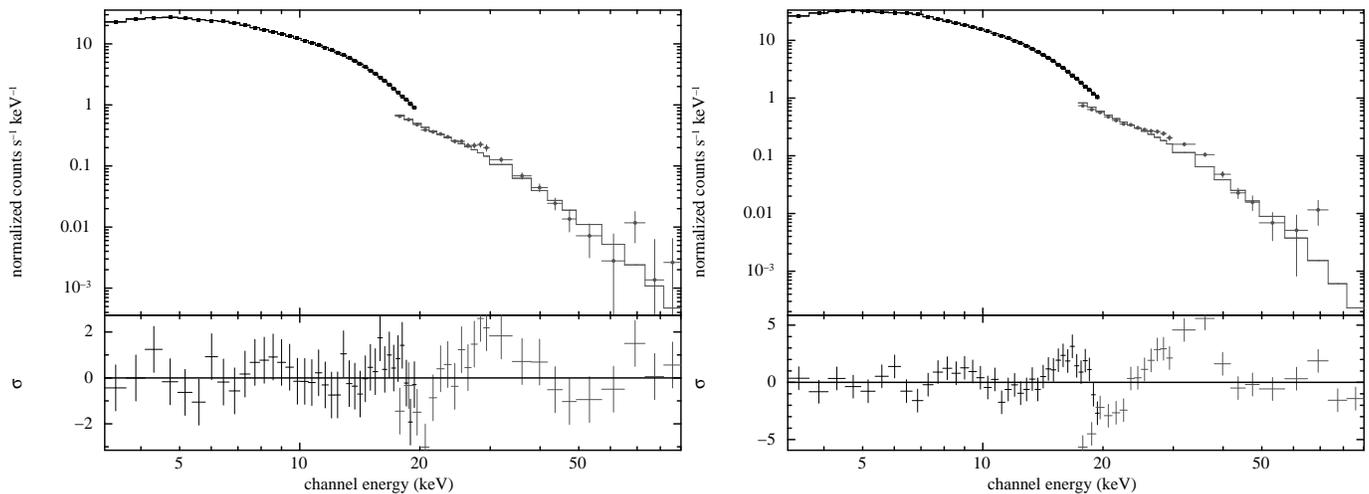

  \centering
  \includegraphics[angle=-90,width=9cm]{12815fg1.ps}
  \includegraphics[angle=-90,width=9cm]{12815fg2.ps}
  \caption{Combined spectrum and model of data obtained
    with \emph{PCA} (3$-$20 keV) and \emph{HEXTE} (17$-$100 keV).
    Both data sets belong to the run carried out in 2001 and 
    their orbital phases are 0.53 and 0.66, respectively.
    Bottom panels show the residuals in units of $\sigma$ with respect to the model
    (see Section~\ref{RXTEanalysis} for details).
  }
  \label{rxte}
\end{figure*}

The pulse-phase averaged X-ray spectrum of 4U 1538$-$52 has usually
been described either by an absorbed power law modified by a high
energy cutoff, a power law modified by a Fermi-Dirac cutoff, or by two
power laws with indices of opposite sign multiplied by an exponential
cutoff (the NPEX model, Mihara~\cite{mihara}; Rodes et al. \cite{rodes1}).
In addition to these continuum models, an iron
fluorescence line at $\sim$6.4 keV and a cyclotron resonant scattering
feature at $\sim$20 keV discovered by \emph{Ginga} (Clark et al.
\cite{clark90}) are needed to describe the data. The variability of
this CRSF was studied by Rodes-Roca et al. (\cite{rodes2}).
Rossi X-ray Timing Explorer (\emph{RXTE}) (Coburn \cite{coburn}) and
\emph{BeppoSAX} data (Robba et al.  \cite{robba}) did not show
evidence of the first harmonic at $\sim$40 keV. Robba et
al. (\cite{robba}) presented some evidence of an absorption feature
around 50 keV; however, because of the lack of a signal-to-noise ratio of the
spectrum at these energies, the feature could not be confirmed.
   
In this paper, we report on the 3--100 keV analysis based on the
observations of 4U 1538$-$52 performed by the \emph{RXTE} and
\emph{INTEGRAL} satellites. In Sect.~\ref{data} we describe the
observations and data analysis. In Sect.~\ref{analyse} spectral
analysis are presented, and summarized in Sect.~\ref{conclusion}.

\section{Observations}
\label{data}

\subsection{RXTE data}

To study the presence of spectral features, we have used all archival
data from \emph{RXTE} on this source: three observations have been
carried out in 1996, 1997, and 2001. The first one, is a monthly
observation between 1996 November 24 and 1997 December 13. The second
and third ones cover a complete orbital period. In our analysis we
used data from both \emph{RXTE} pointing instruments, the Proportional
Counter Array (\emph{PCA}) and the High Energy X-ray Timing Experiment
(\emph{HEXTE}).

To extract the spectra, we used the standard
\emph{RXTE} analysis software FTOOLS\footnote{available at \emph{http://heasarc.gsfc.nasa.gov}}.
This package takes care of the modeling of \emph{PCA} background, the dead
time corrections of \emph{HEXTE} data and generates the appropriate response matrices
for the spectral analysis.
   
The \emph{PCA} consists of five co-aligned Xenon proportional counter
units with a total effective area of $\sim$6000 cm$^2$ and a nominal
energy range from 2 keV to over 60 keV (Jahoda et
al. \cite{jahoda}). However, because of response problems above $\sim$20
keV and the Xenon-K edge around 30 keV, we restricted the use of the
\emph{PCA} to the energy range from 3 keV to 20 keV (see also Kreykenbohm et
al. \cite{ingo}).
Systematic uncertainties
are taken into account by the standard spectral analysis.
   
The \emph{HEXTE} consists of two clusters of four NaI(Tl)/CsI(Na)
Phoswich scintillation detectors with a total net detector area of
1600 cm$^2$.  These detectors are sensitive from 15 keV to 250 keV
(Rotschild et al. \cite{hexte}), however, response matrix, instrument
background and source count rate, limit the energy range from 17 to
100 keV. Background subtraction in \emph{HEXTE} is done by
source-background swapping of the two clusters every 32\,s throughout
the observation.
For the \emph{HEXTE}, the response matrices were generated with HXTRSP,
version 3.1. We used HXTDEAD version 2.0.0 to correct for the dead time.
In order to improve the statistical significance of
the data, we added the data of both \emph{HEXTE} clusters and created
an appropriate response matrix by using a 1:0.75 weighting to account
for the loss of a detector in the second cluster. We also binned
several channels together of the \emph{HEXTE} data at higher energies
and chose the binning as a compromise between increased statistical
significance while retaining a reasonable energy resolution.

\subsection{\emph{INTEGRAL} data}
\label{integral-data}
   
\emph{INTEGRAL} \cite{winkler03} has a unique broad band capability
thanks to its four science instruments (imager \emph{IBIS},
spectrometer \emph{SPI}, X-ray monitor \emph{JEM-X}, and optical
monitor \emph{OMC}) which allow us to study a source from 3\,keV up to
10\,MeV and in the optical simultaneously. The imager \emph{IBIS} has
a very large field of view of $9^\circ \times 9^\circ$ which allows us to
observe many sources at the same time. Together with a large
collecting area of 2600\,cm$^2$ and decent energy resolution of 9\% at
100\,keV makes \emph{IBIS} the prime instrument for our analysis.

Since its launch on October 17 2002, \emph{INTEGRAL} observatory
has been constantly collecting a wealth of data. \emph{INTEGRAL}
data are organized in revolutions (i.e. 72\,h long satellite orbits
around the Earth) and then science windows which are typically 1800\,s
to 3600\,s long.

\begin{figure}[htb]
  \centering
  \includegraphics[width=9cm]{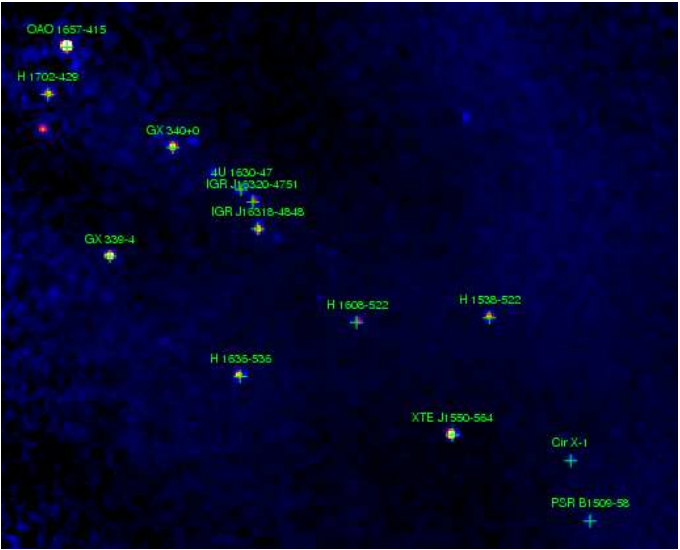}
  \caption{\emph{IBIS/ISGRI} mosaic image of the Norma Arm region where the source
  is located and clearly detected.
  }
  \label{isgri}
\end{figure}

The source was observed by the hard X-ray imager (\emph{IBIS}) camera
on board \emph{INTEGRAL} during the regular scans of the Galactic
plane and the Norma survey. Fig.~\ref{isgri} shows an \emph{IBIS/ISGRI}
mosaic image of the Norma Arm region in the energy range 20$-$100 keV.
The source 4U 1538$-$52 (labelled as H 1538$-$522) is clearly detected and its
position is well determined, allowing us to extract its spectrum
without contamination of other sources in the field of view of the instrument.

To extract \emph{INTEGRAL} data, we used the Offline Science
Analysis Software (OSA) version 7.0, as distributed by the Integral
Science Data Centre (ISDC) following the respective
cookbook\footnote{available at \emph{http://isdc.unige.ch}}
instructions. For \emph{IBIS}, we selected all public data up to
revolution~400 with the source in the fully coded field of view of the
instrument, resulting in $\sim$400 Science Windows (ScWs). In a first
step we created a mosaic of the full data set to obtain a catalog of
the detected sources in the field of view. Since 4U\,1538$-$522 is
relatively close to the galactic centre, the catalog contains other 15
sources, among them several bright sources such as 4U\,1700$-$377. We
then used this catalog to extract the spectrum of 4U\,1538$-$522 using
all 400 ScWs in order to obtain a high signal-to-noise ratio. Since all
other sources in the field are several degrees away, source confusion
is not a problem.  For \emph{JEM-X} we used all data
of monitor number one (\emph{JEM-X 1}) within 2$^\circ$
to ensure a reliable spectrum resulting in approximately 55 ScWs. And
for \emph{SPI}, the selection of all available data within a 8$^\circ$
radius (the fully coded field of view of the instrument) allowed us to obtain
969 ScWs. The selected spectra then have an effective exposure
time of 720\,ksec for \emph{IBIS}, 1907\,ksec for \emph{SPI}, and
360\,ksec for \emph{JEM-X}.

See Table~\ref{obsdata} for the details of all observations which have been used in this
work.

\begin{table}
  \caption{\bf{Details of observations.}}                
\label{obsdata}
\centering                          
\begin{tabular}{l c c c}        
\hline\hline                 
Satellite/Instrument & MJD & Exposure &  Orbital  \\    
 & & (ks) & phase  \\
\hline                        
{\rm RXTE/PCA} & 51925.25 & 9.296 & 0.53 \\      
{\rm RXTE/HEXTE} & 51925.25 & 6.418 & 0.53 \\
{\rm RXTE/PCA} & 51925.74 & 12.592 & 0.66 \\      
{\rm RXTE/HEXTE} & 51925.74 & 8.261 & 0.66 \\
  &  &  & \\
{\rm  INTEGRAL/JEM-X 1} & 53407.7$-$53826.3 & 360 & \\
{\rm  INTEGRAL/ISGRI} & 52651.4$-$53756.2 & 720 & \\
{\rm  INTEGRAL/SPI} & 52651.4$-$53941.4 & 1907 & \\
\hline                                   
\end{tabular}
\end{table}

\begin{figure}[htb]
  \centering
  \includegraphics[angle=0,width=9cm]{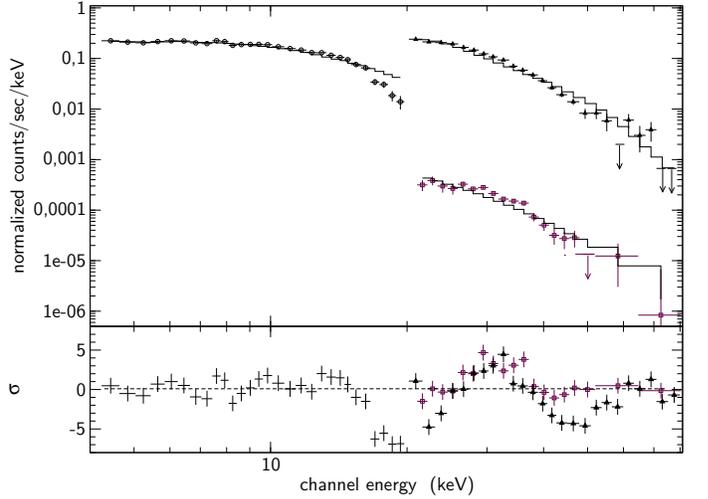}
  \caption{Combined spectrum and model obtained with \emph{JEM-X}
    (\emph{left}), \emph{ISGRI} (\emph{upper right}), and \emph{SPI}
    (\emph{lower right}). The continuum is modeled by the
    \textsc{cutoffpl} model without any cyclotron lines applied.
    The bottom panel shows the residuals in units of $\sigma$
    with
    respect to the model. The first harmonic at $\sim$47 keV is
    already evident in the raw data.}
  \label{integral}
\end{figure}

\section{Spectral analysis}
\label{analyse}

For spectral analysis we used the \textsc{XSPEC} (\cite{arnaud}) fitting package,
released as a part of \textsc{XANADU} in the HEASoft tools.

\subsection{\emph{RXTE} analysis}
\label{RXTEanalysis}

Fig.~\ref{rxte} shows \emph{PCA} and \emph{HEXTE} phase-averaged spectrum at
orbital phases 0.53 (left panel) and 0.66 (right panel). The raw data
together the best-fit model and the residuals of the fit as the difference
between observed flux and model flux divided by the uncertainty of the
observed flux, i. e. in units of $\sigma$,
are included in this plot. The dip
of the second cyclotron line at $\sim$ 47 keV is apparent in the raw data.
The \emph{RXTE} continuum is properly described by an absorbed powerlaw modified
by an exponential cutoff at $E_{\rm cut}=9.4^{+1.1}_{-0.23}$ keV ({\sc
  cutoffpl} in \textsl{XSPEC} (\cite{arnaud})), see Table~\ref{rxteresults}.
 Analytically, \textsc{cutoffpl} is given by the equation,
   \begin{equation}
   cutoffpl(E) = K \, \left( E/1keV \right)^{-\alpha} \, e^{-E/E_{cut}} \,,
   \end{equation}
   where $\alpha$ is the power law photon index and $E_{cut}$ is the cutoff
   energy in keV.
An iron emission line at 6.4\,keV
was also added. The proper description of the continuum, however,
further required a very broad Gaussian emission line
component at 12
keV to account for a flux excess at these energies. This
phenomenological component has been used to describe the continuum of
several X-ray pulsars (Coburn et al.~\cite{coburn2} concluded that
this bump is an inherent feature in the spectra of accreting pulsars).
Although it has no clear physical
interpretation, its use here is justified as we concentrate on the
presence of absorption features at significantly higher
energies. See Table~\ref{rxteresults}
for the best fit spectral parameters for the data set in Fig.~\ref{rxte}.
Data from \emph{PCA} and \emph{HEXTE} have been fitted simultaneously.
All uncertainties refer to a single parameter at the 90\% ($\Delta\chi^2 = $ 2.71)
confidence limit\footnote{Also in Tables \ref{integral}, \ref{isgrispihexte}
and \ref{crsfisgri}}.

A significant absorption feature is present in the residuals around
$\sim$47 keV. When other models are used to describe the underlying
continuum, i.e. an absorbed powerlaw with a Fermi-Dirac cutoff
(\cite{tanaka} 1986) or the
NPEX model (Mihara~\cite{mihara}), the overall residuals were higher,
but none of them was
able to account for the absorption feature at $\sim$47 keV.

\begin{table}
  \caption{Fitted parameters for the \emph{RXTE} spectra in Fig.~\ref{rxte}.
    }
\label{rxteresults}
\centering                          
\begin{tabular}{r r c c}        
\hline\hline                 
Component & Parameter & Left panel & Right panel \\    
\hline                        
{\rm Continuum} &   $\alpha$ & 0.74$^{+0.18}_{-0.14}$ & 0.52$\pm 0.11$ \\      
  & $E_{cut}$ (keV) & 11.4$^{+2.0}_{-1.2}$ & 9.4$^{+1.1}_{-0.23}$ \\
  & $N_H$ (10$^{22}$ cm$^{-2}$) & 1.0$^{+0.6}_{-0.5}$ & 1.05$^{+0.10}_{-0.20}$ \\
  &  &  & \\
{\rm Gaussian}  & $E_{bump}$ (keV) & 12.54$^{+0.21}_{-0.3}$ & 12.15$^{+0.10}_{-0.09}$ \\
{\rm emission line}  & $\sigma_{bump}$ (keV) & 3.12$^{+0.22}_{-0.23}$ & 2.92$^{+0.08}_{-0.12}$ \\
  &  &  & \\
{\rm  Fluorescence} & $E_{FeK}$ (keV) & 6.54$^{+0.14}_{-0.11}$ & 6.56$^{+0.03}_{-0.14}$ \\
{\rm  iron line}   & $\sigma_{FeK}$ (keV) & 0.2$^{+0.3}_{-0.19}$ & 0.1 (frozen) \\
 & & & \\
 &  $\chi^2_{\nu}$(dof) & 1.2(51) & 4.4(52) \\ 
\hline                                   
\end{tabular}
\end{table}

\subsection{\emph{INTEGRAL} analysis}
\label{INTEGRALanalysis}

In Fig.~\ref{integral}, we present the combined \emph{JEM-X},
\emph{ISGRI} and \emph{SPI} spectrum using all data. The spectra of
all three instruments were fitted simultaneously with a
\textsc{cutoffpl} model, using the latest response matrices
provided by the OSA software (see Section~\ref{integral-data} for more details).
A factor term was included in the model to
allow for the adjustement of efficiencies between different instruments. No
photoelectric absorption has been used since the {\it INTEGRAL}
spectra start only at 4 keV and none was required by the data. An
absorption column below $10^{23}$ cm$^{-2}$ is expected to have no
noticeable effect below 5 keV. 
Nevertheless \emph{JEM-X} data are needed to constrain the X-ray continuum
model.
Likewise, neither the very broad
Gaussian at 12 keV, nor the iron fluorescence line at 6.4\,keV are
required by the \emph{JEM-X} data and were therefore not added to the
model. As can clearly be seen in Fig.~\ref{integral}, significant
absorption line structures are present at $\sim$20\,keV (the well
known CRSF) and $\sim47$\,keV
after fitting the continuum with a cutoffpl model (best-fit parameters
of continuum are given in Table~\ref{integral-fit}). Data from
\emph{JEM-X}, \emph{ISGRI} and \emph{SPI} have been fitted simultaneously.

\begin{table}
  \caption{Fitted parameters for the \emph{INTEGRAL} spectrum
  in Fig.~\ref{integral}.
    }
\label{integral-fit}
\centering                          
\begin{tabular}{r r c c}        
\hline\hline                 
Component & Parameter &  &    \\    
\hline                        
{\rm Continuum} &   $\alpha$ & & 0.29$\pm 0.07$ \\      
  & $E_{cut}$ & (keV) & 9.06$^{+0.22}_{-0.21}$ \\
  &  &  & \\
{\rm Normalization}  & $C_{JEM-X}$ & & 0.90$\pm 0.03$ \\
  & $C_{ISGRI}$ & & 1.0 (frozen) \\
  & $C_{SPI}$ & & 1.29 $^{+0.08}_{-0.07}$ \\
  &  &  & \\
 &  $\chi^2_{\nu}$(dof) &  & 5.5(68) \\ 
\hline                                   
\end{tabular}
\end{table}

\subsection{Combined \emph{RXTE} and \emph{INTEGRAL} analysis}
\label{combined-analysis}

To further explore this second feature and in order to achieve the
highest significance at high energies, we jointly fitted the data of
the high energy instruments of the two satellites, \emph{HEXTE},
\emph{ISGRI}, and \emph{SPI}. We fixed the continuum parameters
$\alpha$, the power law photon index and $E_{cut}$, the cutoff energy
of the exponential cutoff, to the values given
in Table~\ref{integral-fit}.

\begin{figure}[h!t]
  \centering
  \includegraphics[width=9cm]{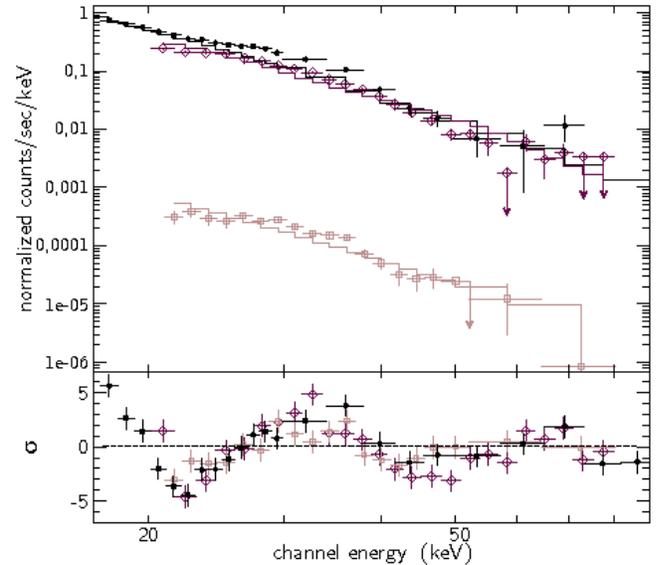}
  \caption{Spectra and continuum model (cutoffpl)
    obtained with \emph{ISGRI}, \emph{SPI} (\emph{lower}), and
    \emph{HEXTE}. The bottom panel shows the residuals in units of
    $\sigma$ of a fit without including a cyclotron line.
    }
  \label{newcrsf}
\end{figure}

In Table~\ref{isgrispihexte} we show the best fit parameters for each
spectrum we used in Fig.~\ref{newcrsf}. Data from
\emph{SPI}, \emph{ISGRI} and \emph{HEXTE} have been fitted individually.
As the continuum parameters are frozen, we have not included them in this Table.
The results of the spectral fits
from three different instruments show that the values describing the \emph{shape}
of the cyclotron lines are consistent with one to another within uncertainties.
  
\begin{table}
  \caption{Fitted parameters for the fundamental and the first harmonic CRSF.
    }
\label{isgrispihexte}
\centering                          
\begin{tabular}{l c c c}        
\hline\hline                 
Parameter & \emph{SPI} & \emph{ISGRI} & \emph{HEXTE}   \\    
\hline                        
{\rm  Fundamental} &  &  &  \\
$E_{c_1}$ (keV) & 22.7$^{+0.9}_{-2.4}$ & 22.0$^{+1.0}_{-1.6}$ & 20.7$^{+1.3}_{-3}$ \\
$\tau_{c_1}$  & 0.8$^{+0.4}_{-0.3}$ & 0.5$^{+1.6}_{-0.18}$ & 0.5$^{+1.3}_{-0.13}$ \\
$\sigma_{c_1}$ (keV) & 3$^{+4}_{-1.7}$ & 3$^{+5}_{-2}$ & 4$^{+6}_{-3}$ \\
{\rm $1^{st}$ Harmonic} &  &  &  \\
$E_{c_2}$ (keV) & 42$^{+4}_{-1.5}$ & 47$^{+3}_{-10}$ & 48$^{+9}_{-7}$ \\
$\tau_{c_2}$  & 0.9$^{+1.0}_{-0.5}$ & 0.8$^{+0.5}_{-0.21}$ & 0.6$^{+1.7}_{-0.3}$ \\
$\sigma_{c_2}$ (keV) & 3$^{+11}_{-1.9}$ & 10$^{+22}_{-6}$ & 8$^{+24}_{-7}$ \\
 & & & \\
$\chi^2_{\nu}$(dof) & 1.0(14) & 1.1(18) & 0.8(16) \\ 
\hline                                   
\end{tabular}
\end{table}

\begin{table}
  \caption{Best fit parameters for the fundamental and the first harmonic CRSF.
    }
\label{crsfisgri}
\centering                          
\begin{tabular}{r r c c}        
\hline\hline                 
Component & Parameter &  &    \\    
\hline                        
{\rm Continuum} &   $\alpha$ & & 0.29 (frozen) \\      
  & $E_{cut}$ & (keV) & 9.1 (frozen) \\
  &  &  & \\
{\rm  Fundamental} & $E_{c_1}$ & (keV) & 21.4$^{+0.9}_{-2.4}$ \\
   & $\tau_{c_1}$ &  & 0.43$^{+0.3}_{-0.07}$ \\
   & $\sigma_{c_1}$ & (keV) & 3$^{+4}_{-2}$ \\
 & & & \\
{\rm $1^{st}$ Harmonic} &    $E_{c_2}$ & (keV) & 47.1$^{+2.2}_{-1.7}$ \\
   & $\tau_{c_2}$ &  & 0.88$^{+0.5}_{-0.24}$ \\
   & $\sigma_{c_2}$ & (keV) & 5$^{+5}_{-2.3}$ \\
 & & & \\
 &  $\chi^2_{\nu}$(dof) &  & 1.4(41) \\ 
\hline                                   
\end{tabular}
\end{table}

\begin{figure}[h!t]
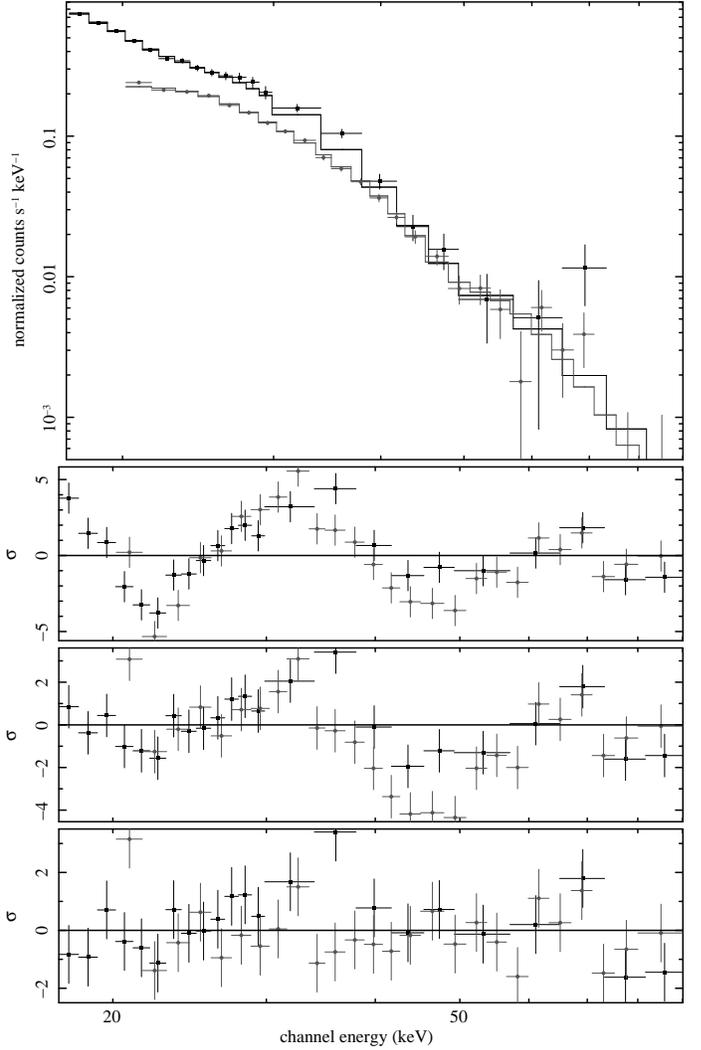

  \centering
  \includegraphics[angle=-90,width=9cm]{12815fg6.ps}
  \includegraphics[angle=-90,width=9cm]{12815fg7.ps}
  \includegraphics[angle=-90,width=9cm]{12815fg8.ps}
  \includegraphics[angle=-90,width=9cm]{12815fg9.ps}
  \caption{Cyclotron line modeling of the phase-averaged spectra
    \emph{HEXTE/ISGRI} of 4U 1538$-$52.
    \emph{Top panel}: Spectra and best fit model (cutoffpl and two cyclotron lines)
    obtained with \emph{HEXTE} and
    \emph{ISGRI}. Bottom panels show the residuals in units of
    $\sigma$ for models taking different numbers of cyclotron lines into account.
    \emph{Second panel}: Without cyclotron lines.
    \emph{Third panel}: Fundamental cyclotron line at $\sim$22 keV.
    \emph{Bottom panel}: Two cyclotron lines, the fundamental at $\sim$22 keV
    and the first harmonic $\sim$47 keV (see Table~\ref{crsfisgri} for the
    best fit values).
    }
  \label{ftest}
\end{figure}

As evident from Fig.~\ref{newcrsf}, the two cyclotron
absorption features are clearly seen in the raw data of the three
instruments. Although
\textsl{SPI} has a very high energy resolution, the compared to
\textsl{ISGRI} rather low signal-to-noise ratio required the use of
rather broad energy bins.  Therefore we will use \textsl{ISGRI} as the
prime instrument in our study together with \textsl{HEXTE} while
\textsl{SPI} data will be used for comparison. In summary, to get the
best fit parameters of the absorption features, we have combined
\textsl{ISGRI} and \textsl{HEXTE} data including a factor for free
normalization between the two instruments (see Fig. ~\ref{ftest} top panel).
The two absorption features at $\sim$22 keV and $\sim$47 keV
are modeled using the {\sc cyclabs} model from {\sc XSPEC} (\cite{arnaud}).

We started by modeling the fundamental cyclotron line that was
discovered by the \emph{Ginga} satellite observatory (Clark et al.~\cite{clark90})
(see Fig.~\ref{ftest} second panel). After the
inclusion of the fundamental cyclotron line at $E\sim 21.4$ keV, the
$\chi^2_{\nu}$ improves from 5.5 for 47 degrees of freedom (dof) to
3.4, for 44 dof (F-test: $3.0\times 10^{-5}$).
Although residuals improve significantly
(see Fig.~\ref{ftest} third panel),
another absorption feature can be seen at $\sim$47 keV.
Including a second
cyclotron line at $\sim47$ keV improves the fit further resulting in a
$\chi^2_{\nu}$ of 1.4 for 41 dof (F-test: $1.7\times 10^{-8}$
see Fig.~\ref{ftest} bottom panel).
\emph{ISGRI} data (filled grey circles in Fig.~\ref{ftest})
has a far better resolution than \emph{HEXTE} at these energies. However,
the overall shape is virtually identical, ruling out any instrumental or
circumstantial effects.
The final fit parameters are given in Table~\ref{crsfisgri}.
Data from \emph{HEXTE} and \emph{ISGRI} have been fitted simultaneously.

The F-test is known to be problematic when used to
test the significance of an additional spectral feature (see Protassov
\& van Dik \cite{ftest}), even if systematic uncertainties are not an
issue. However, the low false alarm probabilities may make the
detection of the line stable against even crude mistakes in the
computation of the significance (Kreykenbohm \cite{ingophd}).
Therefore, taking into account these caveats, we can conclude that the
the first harmonic CRSF is detected with high significance in the
spectrum of this source.

Although the uncertainties of the determined values are rather large,
these values are consistent with one another within uncertainties.
This makes it extremely unlikely that the $\sim$47 keV
feature results from a calibration problem.
Nevertheless, the line
parameters depend slightly on the shape of the continuum (for example,
using the NPEX component (Mihara~\cite{mihara}) and two cyclotron absorption lines,
we obtained the line centre energies at $\sim$22 keV and $\sim$44 keV,
for the fundamental and first harmonic respectively).

In short, this feature has been found to be present under the following
circumstances:
\begin{itemize}
\item Three different telescopes and instruments:  Narrow Field Instruments (\emph{NFIs})
  on board \emph{BeppoSAX} (Robba et al. \cite{robba}), \emph{HEXTE} on board
  \emph{RXTE}, and \emph{IBIS/ISGRI} on board \emph{INTEGRAL} (this work).
\item Different epochs within the same telescope (i.e. 1996, 1997, 2001 using \emph{RXTE}).
\item Different orbital phases within a given epoch and instrument
  (Fig.~\ref{rxte}) .
\end{itemize}
We therefore conclude, that despite the relatively low significance of
the feature in an \emph{individual} observation, its detection in
different epochs, configuration of the binary system, and different
instrumentation, makes this feature highly significant. The most
direct explanation for this absorption line like feature is that it is
the first harmonic cyclotron line.

\section{Summary and discussion}
\label{conclusion}

We presented the spectral analysis of 4U 1538$-$52 using data from
\emph{RXTE} and \emph{INTEGRAL}. We present evidence for a previously
unknown absorption line like feature in the phase-averaged spectrum of
the source.  As we have shown in Section~\ref{analyse}, we have been
able to achieve a good fit to the phase-averaged spectra by including
a Lorentzian absorption line at $\sim$47 keV into the model (see
Fig.~\ref{ftest}). This absorption line is clearly
visible whenever the signal-to-noise ratio in the spectrum is good
enough to allow an analysis of the data. The most straightforward
interpretation for this feature is that it is the first harmonic of
the $\sim$ 22\,keV fundamental CRSF.
   
According to the theory, cyclotron lines are due to the resonant
scattering of photons by electrons whose energies are quantized into
Landau levels by the magnetic field (M\'esz\'aros
\cite{meszaros}). The quantized energy levels of the electrons are
harmonically spaced in the first order, such that the first harmonic
line should be placed at twice the energy of the fundamental line,
i.e. $ 2\times E_{\rm cyc}\approx 42$ keV. In reality, however, the
coupling factor between the fundamental and first harmonic is with
$\sim 2.20$ slightly higher than 2.0.  This anharmonic spacing,
however, has been observed already in several systems where more than
one line is present. As explained by \cite{schonherr07} (2007), the
relativistic photon-electron scattering already produces some
anharmonicity, because photons with energies close to the Landau
levels may not escape the plasma if their energies are not changed by
inelastic scattering. This, however, can not be the only reason
because some systems show an anharmonic spacing larger than that
predicted by this effect. A possibility to explain this difference is
to take into account that the optical depths of the fundamental and
the first harmonic could be different if they are formed at different
heights above the neutron star. With increasing height, the strength
of the magnetic field decreases resulting in a different CRSF
energy. Another possibility is to consider a displacement of the
magnetic dipole which would also explain the difference of energy of
the two lines if the lines originate from the different poles of the
neutron star. Therefore a significant phase dependence of the strength
of the both lines is expected, however, the low signal-to-noise ratio at
higher energies prevents us to test this hypothesis with the current
data sets.

\begin{acknowledgements}
  We are grateful to the anonymous referee for
  useful and detailed comments.
  Part of this work was supported by the Spanish Ministry of Education
  and Science \emph{Primera ciencia con el GTC: La astronom\'{\i}a
    espa\~nola en vanguardia de la astronom\'{\i}a europea} CSD200670
  and \emph{Multiplicidad y evoluci\'on de estrellas masivas} project
  number AYA200806166C0303. This research has made use of data
  obtained through the High Energy Astrophysics Science Archive
  Research Center Online Service, provided by the NASA/Goddard Space
  Flight Center and through the INTEGRAL Science Data Center (ISDC),
  Versoix, Switzerland. SMN is a researcher of the Programme Juan de
  la Cierva, funded by the MICINN. JMT acknowledges the support by the
  Spanish Ministerio de Educaci\'on y Ciencia (MEC) under grant
  PR2007-0176. ACA thanks for the support of this project to the
  Spanish Ministerio de Ciencia e Innovaci\'on through the 2008
  postdoctoral program MICINN/Fulbright under grant 2008-0116.
  JJRR acknowledges the support by the
  Spanish Ministerio de Educaci\'on y Ciencia (MEC) under grant
  PR2009-0455.
\end{acknowledgements}

\end{document}